\documentclass[chicago]{emulateapj}
\usepackage{amsmath,epsfig,url,natbib}
\usepackage{graphicx,epstopdf,float,color,array}
\usepackage{rotating,multirow}
\submitted{}
\shorttitle{Deblending with GAN}
\shortauthors{Hemmati et al.}

\definecolor{orange}{cmyk}{0,0.5,1,0}

\begin{document}
\title{Deblending Galaxies with Generative Adversarial Networks}

\author{Shoubaneh Hemmati\altaffilmark{1,2}, Eric Huff\altaffilmark{1}, Hooshang Nayyeri\altaffilmark{3}, Agn\`es Fert\'e\altaffilmark{1}, Peter Melchior\altaffilmark{4}, Bahram Mobasher\altaffilmark{5}, Jason Rhodes\altaffilmark{1}, Abtin Shahidi\altaffilmark{5}, Harry Teplitz\altaffilmark{2}}
\email{shemmati@caltech.edu}

\altaffiltext{1}{Jet Propulsion Laboratory, California Institute of Technology, Pasadena, CA 91109, USA}
\altaffiltext{2}{Infrared Processing and Analysis Center, California Institute of Technology, Pasadena, CA 01109, USA}
\altaffiltext{3}{University of California Irvine, Irvine, CA 92697, USA}
\altaffiltext{4}{Princeton University, Princeton, NJ 08544, USA}
\altaffiltext{5}{University of California Riverside, Riverside, CA 92521, USA}
\journalinfo{\textsuperscript{\textcopyright} 2022. All rights reserved. Accepted for publication in the Astrophysical Journal.}

\begin{abstract}

Deep generative models including generative adversarial networks (GANs) are powerful unsupervised tools in learning the distributions of data sets. 
Building a simple GAN architecture in PyTorch and training on the CANDELS data set, we generate galaxy images with the Hubble Space Telescope resolution starting from a noise vector. We proceed by modifying the GAN architecture to improve the Subaru Hyper Suprime-Cam ground-based images by increasing their resolution to the HST resolution. We use the super resolution GAN on a large sample of blended galaxies which we create using CANDELS cutouts. In our simulated blend sample, $\sim 20 \%$ would unrecognizably be blended even in the HST resolution cutouts. In the HSC-like cutouts this fraction rises to $\sim 90\%$. With our modified GAN we can lower this value to $\sim 50\%$. We quantify the blending fraction in the high, low and GAN resolutions over the whole manifold of angular separation, flux ratios, sizes and redshift difference between the two blended objects. The two peaks found by the GAN deblender result in ten times improvement in the photometry measurement of the blended objects. Modifying the architecture of the GAN, we also train a Multi-wavelength GAN with seven band optical+NIR HST cutouts. This multi-wavelength GAN improves the fraction of detected blends by another $\sim 10\%$ compared to the single-band GAN.  This is most beneficial to the current and future precision cosmology experiments (e.g., LSST, SPHEREx, Euclid, Roman), specifically those relying on weak gravitational lensing, where blending is a major source of systematic error.
\end{abstract}


\section{Introduction}

 Observational astronomy has been recording galaxy data for nearly a hundred years. These data inform us, generally, about the variety of existing galaxies, their physical properties, and how they transform and evolve over cosmic time. Equally important, observations of galaxies and their distributions provide an exceptional pool for measuring cosmological parameters and testing for the pillars of modern physics (e.g., gravity). 
 
 With time and technology advancements, we can cover an ever larger volume of the sky (i.e., wider areas and greater depths), a wider range of the electromagnetic spectrum, as well as higher spectral and spatial resolutions. Current and future ground-based and space observatories all help push the boundaries in one or more of these avenues to obtain a better characterization of galaxies and hence a higher precision in cosmological measurements. For instance NASA's James Webb Space Telescope (JWST; \citealt{Gardner2006}), complements and extends Hubble Space Telescope (HST) observations in the infrared. The Euclid mission (\citealt{euclid}) as well as NASA's Nancy Grace Roman Space Telescope (Roman; \citealt{Spergel2015}) with a field of view one hundred times that of the HST will cover a much wider area of the infrared sky. And, the Spectro-Photometer for the History of the Universe, Epoch of Reionization and Ices Explorer (SPHEREx; \citealt{Dore2014}) mission will provide observations in $\sim 100$ photometric bands over the whole sky with low spatial resolution.
 
Quantifying the increase in the information content by each of these new data sets certainly depends on the specific science question in hand. However, given the cost, getting data of every single object in the sky with highest spectral and spatial resolution is not required/optimal. In the big data era, once distributions are well represented, for data points with limited observations, information from the parent distribution could be used as a prior to generate enhanced data products for those objects. 
 
 Generative models are unsupervised machine learning methods which learn the probability distribution from training data and generate new data instance based on the learned distribution (e.g., \citealt{generative}). Domain of data can vary from images and sound waves, to words. For instance, algorithms predicting the next most probable word in a sequence, or creating a new dog image based on learning from other dog images are examples of generative models. 
\begin{figure*}
\centering
  \includegraphics[trim=0cm 0cm 0cm 0cm, clip,width=1.0 \textwidth] {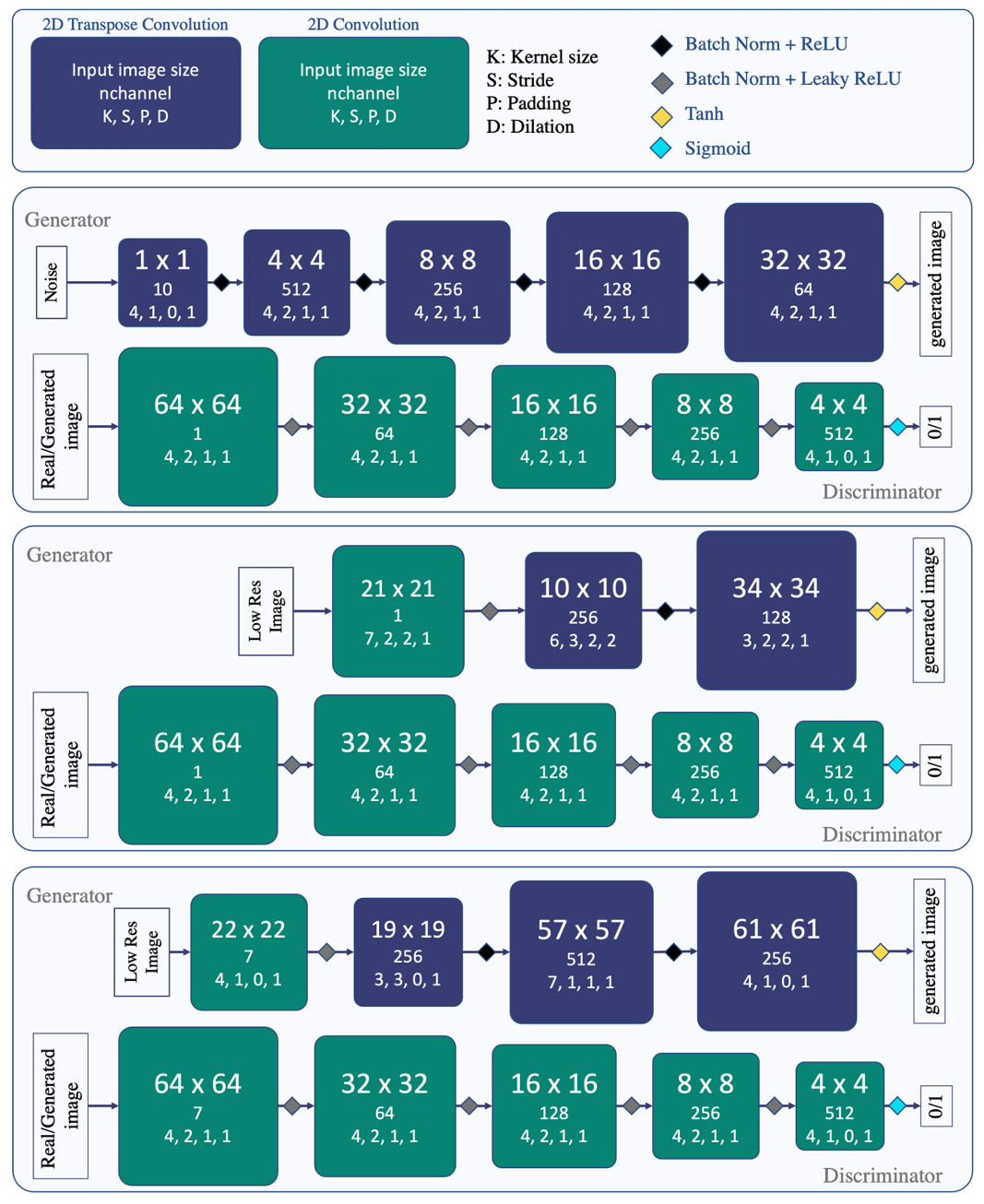}
\caption{The architecture diagrams of the GAN frameworks used in this study. Figure legend shown on the top panel, specifies the symbols used in the convolution, normalization, and activation layers. The second panel corresponds to the vanilla GAN where the input to the generator is a noise vector. The third and fourth panels show the conditional single and multi-waveband GANs with the low resolution and downgraded input images.}
\label{fig:architecture}
\end{figure*}

Generative Adverserial Networks (GANs; \citealt{Goodfellow2014}), inspired by game theory, are generative models which simultaneously tune two competing deep networks; a {\it generator} and a {\it discriminator}. The generator can get as input a 1-dimensional noise array and converts it to an image. The discriminator gets the generated image as well as real images from the training sample and finds the probability of the generated image being drawn from the real distribution. Initially, the discriminator will outperform the generator but with time as the process moves forward, generator gets better in generating data similar to the training distribution, and the discriminator gets better in distinguishing real from fake data. Once the discriminator can no longer distinguish the real and generated images, we stop the training and use the trained generator as an stand alone network.

GANs are extensively used in data-rich regimes and modified architectures are constantly developed in computer science disciplines. Some examples are SPA-GAN for image to image translation (\citealt{Emami2019}); StackGAN which translates text to image (\citealt{Zhang2017}); StyleGAN for generating faces with specified styles (\citealt{Karras2018}). Therefore, besides the mentioned application of GANs to generate images similar to the training distribution from a noise vector, the technique can also be adopted  to increase the resolution of images (i.e., super resolution), to fill in missing parts of images, or to generate more precise images using labels or images in other domains. These functionalities, when applied to astronomy, open new windows and possibilities. 

A variant of GANs have recently been used to simulate realistic looking local galaxy images (\citealt{Dia2019}). \cite{Schawinski2017} used GANs on local galaxies from the Sloan Digital Sky Survey (SDSS; \citealt{SDSS}) to recover structure beyond the deconvolution limit. \cite{Lanusse2019} proposed a hybrid combination of analytic forward modeling of likelihood with deep neural network models, using PixelCNN++ as an optimum deblending approach. \cite{Arcelin2020} also showed the performance of variational auto encoders (VAEs; another class of generative models) in deblending modeled galaxies from surveys such as the Rubin Observatory Legacy Survey of Space and Time (LSST) to be promising. 

\begin{figure*}
\centering
  \includegraphics[trim=0cm 0cm 0cm 0cm, clip,width=1.0 \textwidth] {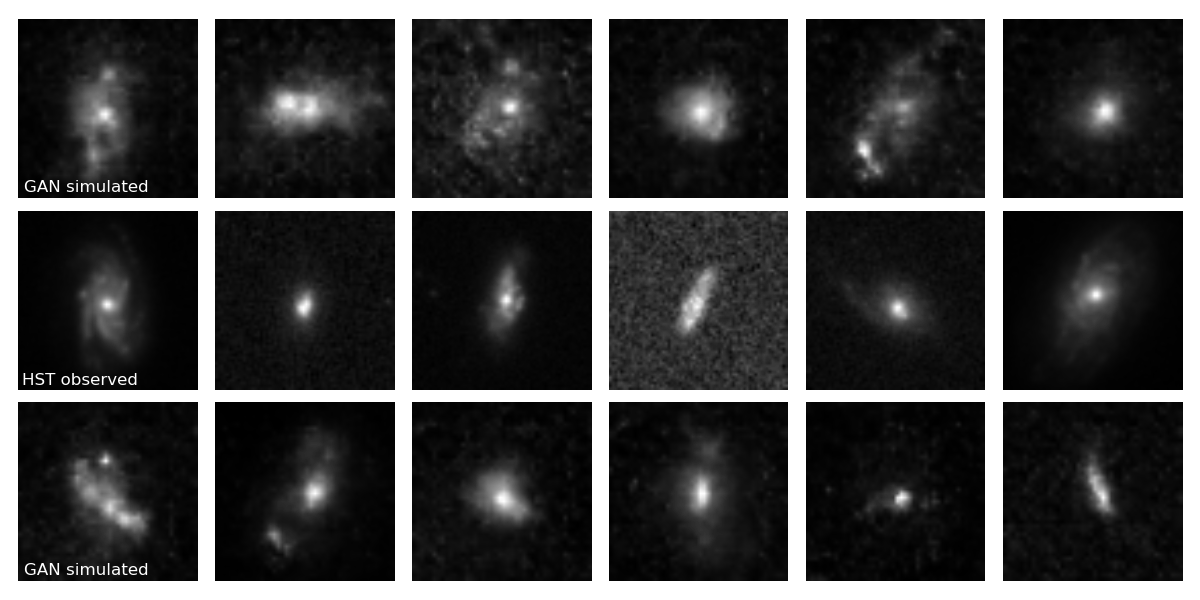}
\caption{GAN with a simple architecture and no input. We trained the network with three thousand sample cutouts in the HST ACS F775W centered on GOODS south galaxies. First and third rows are generated with GAN and the second row shows examples from the training set.}
\label{fig:noise}
\end{figure*}

 \begin{figure*} 
\centering
  \includegraphics[trim=0cm 0cm 0cm 0cm, clip,width=1.0 \textwidth] {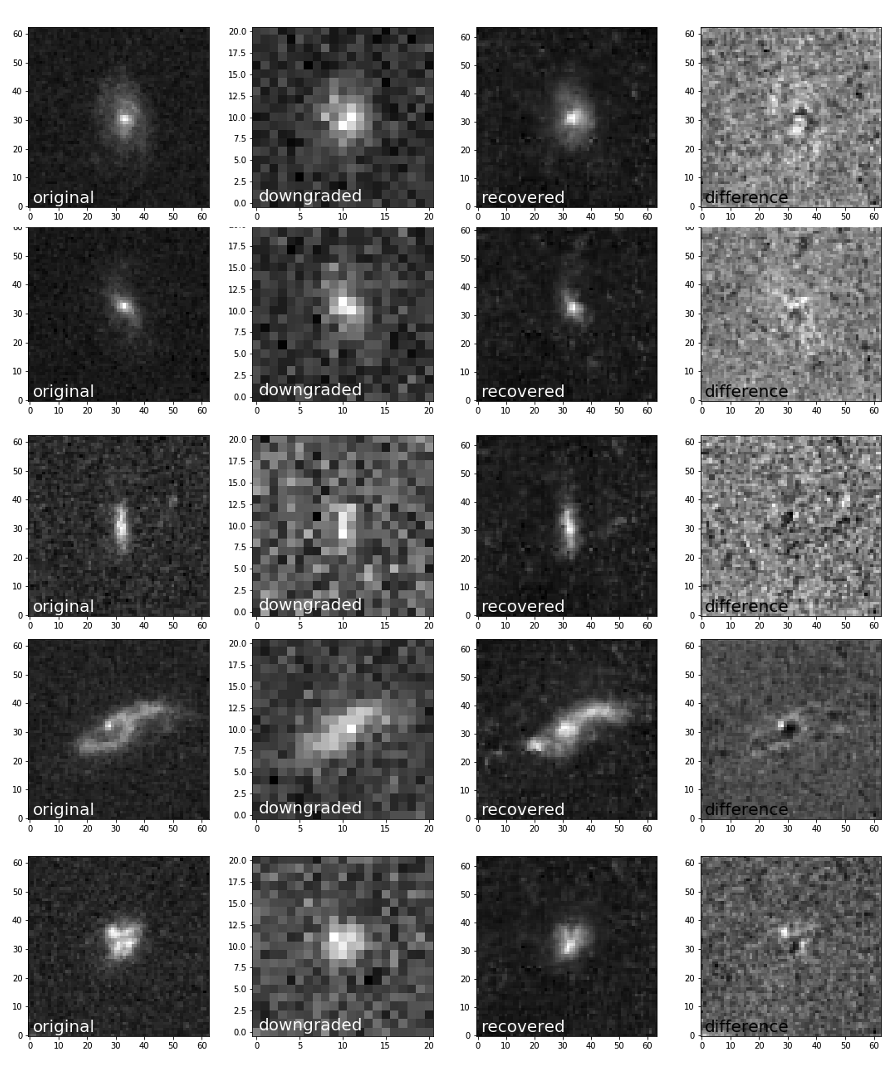}
\caption{Five examples of GAN improving resolution of cutouts from Subaru HSC resolution to HST. From left to right, first Column shows high resolution HST F775W cutouts. Downgraded image to Subaru HSC resolution and pixel scale which is fed to GAN is shown on the second column. Third column shows GAN output improving the resolution of the downgraded image. Difference between the first and third columns are shown in the fourth column.}
\label{fig:resolution}
\end{figure*}
In this paper, we explore the parameter space where the spatial resolution improvement by GANs helps with galaxy blending detection in low resolution images. We introduce the training and test samples in \S \ref{sec:data}. In \S \ref{sec:gan}, we present the architecture of the deep neural networks followed by the results in \S \ref{sec:results}. Finally in \S \ref{sec:disc}, we discuss ways to build upon this machinery for future large surveys. Throughout this paper all magnitudes are expressed in AB system (\citealt{Oke1983}).


\section{Data}\label{sec:data}

In this study, we use the HST Advance Camera for Survey (ACS) images in the F775W from the Cosmic Assembly Near-infrared Deep Extragalactic Legacy Survey (CANDELS; \citealt{Grogin2011}, \citealt{Koekemoer2011}) in the GOODS South field as our high resolution sample. The images have $0.06 \arcsec$ pixel scale, $0.08 \arcsec$ full width at half maximum (FWHM) resolution, $28.5$ (5$\sigma$) depth, and the effective wavelength of the filter is at 7693 \AA. The sample is selected from the CANDELS GOODS South catalog (\citealt{Guo2013}) and is cut in half by declination for training and testing sets. Sources fainter than 25 magnitude in F160W, those with the \textit{CLASS STAR} parameter larger than 0.95 in the CANDELS catalog (measured by the Source Extractor software; \cite{Sextractor}), and galaxies with redshifts higher than five are not included in the training sample. We create 64x64 pixels cutouts centered on the RA and DEC of objects and take the inverse hyperbolic sine function of fluxes (see \citealt{Lupton1999}) before using them in network training. 

The lower resolution sample is made by degrading the HST cutout's FWHM resolution (to 0.6 $\arcsec$), pixel scale (to 0.18 $\arcsec$) as well as adding noise to them. This makes the low resolution sample very similar to those observed by the Subaru Hyper Suprime Cam (HSC; \citealt{Aihara2018}) in i-band (with the effective wavelength of 7659 \AA). The change in resolution is performed using a two dimensional convolution with a matching kernel created from the HST and HSC i-band PSFs. This downgrading scheme is chosen as Subaru has observed a very large area in the sky (1400 $deg^{2}$) and given the ability of GANs in increasing the resolution one can potentially extend HST observations within the CANDELS fields where only a small area is covered ($\sim 0.2$  $deg^{2}$) to thousands of square degrees.

We also build an artificially blended sample, both in high and low resolution, by stacking two or more galaxy cutouts and recording their positions, redshifts, and brightnesses. To do this we follow the same steps as creating the high resolution cutouts but shift and rotate each of cutout randomly before adding them together. Degrading to low resolution is the same as above.

\section{Network Architectures}\label{sec:gan}

We build and train our deep neural networks with PyTorch (\citealt{pytorch}), a machine learning library that performs immediate execution of dynamic tensor computations with automatic differentiation. We use a machine with three Nvidia Titan Xp GPUs each with more than three thousand Cuda cores (\citealt{cuda}) for accelerated computing. In this cluster, it takes $\sim 4$ hours to train a GAN with thousands of galaxy images, compared to days on a CPU machine. In applications with more diverse imaging, where a much larger sample of images are usually needed for training ($> 100,000$), techniques such as adaptive discriminator augmentation (\citealt{Karras2020ada}) can cut the needed sample by 10-20 times, resulting in similar training time for GANs (\citealt{Zhong2020}).

Here, we construct two GAN frameworks (Figure \ref{fig:architecture}). First we show that a vanilla GAN structure with a noise vector as input can learn the distribution of galaxies in the training sample and generate visually realistic galaxy images. And then proceed to the main contribution in which we feed the generator of the GAN with lower resolution sample image and the discriminator compares the generated image to a high resolution one so that the generator learns how to improve the resolution. 

The network architecture of both the generator and the discriminator consists of several layers of convolution. Convolutional neural networks (CNNs; see \citealt{gu2017} for review of advances in CNNs) are well-known deep learning architectures inspired by the visual perception mechanism of living systems. They are now widely used in a variety of problems, such as visual recognition, speech recognition and natural language processing. Each node on the matrix of convolution layers, similar to neurons on a simple vectored-layer neural networks, receives one or more inputs, and then multiplies them by the cell's weights and outputs the sum over them. 

The discriminator network consists of several convolution layers which decreases the input matrix size while the generator consists of layers of transpose convolution which increases the input matrix size. We use the \textit{conv2D} and \textit{convtranspose2D} of Pytorch to define the convolution layers. Each layer is then defined by a set of  hyperparameters, the number of channels in the input and output image, the convolving kernel size, stride of the convolution (number of pixels the filter shifts by over the matrix), padding (number of pixels added to the edges before convolution), and dilation (the spacing between the kernel elements). The main trick in building the network architecture is setting these numbers for all layers in an optimized way where not only the input/output sizes match but also features in images are learned well with a span over a range of kernel sizes. The output of each layer is passed through a batch normalization to stabilize the learning process and activation function layer before getting to the next layer.

Activation functions in artificial neural networks add the non-linearity to the output of a neuron given an input or a set of inputs. We set the activation functions of the neurons in our network, except for the output layers, to ReLU (i.e., Rectifier Linear Unit, \citealt{Relu}) in the generator, and to Leaky ReLU (\citealt{Maas13}) in the discriminator networks. Rectifier functions defined as the positive part of their argument are the most popular activation functions in neural networks. A leaky ReLU allows for a small positive gradient for the non-positive arguments. A hyperbolic tangent and a Sigmoid activation function are used in the last layer of the Generator and discriminator networks respectively (see \citealt{LeCun2012} for detailed differences). The symmetry of a Tanh or Sigmoid in the output layer help with the symmetric treating of darker and lighter colors in images. Our choices of activation functions here are similar to those suggested by \cite{radford2016} in the deep convolutional GAN (DCGAN), who observed that using a bounded activation allowed the model to learn more quickly and to cover the color space of the training distribution. This is in contrast to the initial GAN studies, which used the maxout activation (\citealt{Goodfellow2014}).
 
 We adopt a minimax loss function as suggested by \cite{Goodfellow2014} which uses cross entropy between the real and generated distributions. This loss function is sufficient for the purpose of this study. However more recent literature suggests variations of the GAN loss function (e.g., a Wasserstein loss) can mitigate issues such as training instability or failure to converge (e.g., \citealt{Arjovsky2017}, \citealt{Weng2019}). We use the Adam (derived from adaptive moment estimation) optimizer (\citealt{adam}) to compute the adaptive learning rates which optimize the cost functions of both the generator and the discriminator networks. Adam optimizer is an efficient stochastic gradient decent algorithm with low memory requirements.  
 
Figure \ref{fig:noise} represents sample galaxies made with the first framework (noise vector input) in the first and third rows. The second row shows real GOODS-S galaxy cut-outs in the F750W for comparison. While an expert eye might spot slight differences, many features are learned and visible in the generated images that are often missing in simplified galaxy models such as warped disks, clumps, spiral structures, etc.. Figure \ref{fig:resolution} shows five examples of the modified GAN to improve the resolution. Columns from left to right show the input high resolution (HST ACS) image, the downgraded image (HSC i-band) as the input to the GAN, GAN generated output, and the difference of the generated and original images. We note again that these galaxies are selected from the test sample and were not used in the training of the GAN.

\section{results}\label{sec:results}

Given GAN's ability to efficiently improve the resolution of galaxy images (as seen in Figure \ref{fig:resolution}), they can be employed in astrophysics areas where a higher resolution is favorable. One obvious such area is deblending galaxies with small angular separations. Blending is a major issue in many ground-based surveys; it affects the number counts, photometries, measured shapes and other parameters (see \citealt{Dawson2016}, \citealt{Samuroff2018}, \citealt{Mandelbaum2018}, \citealt{Melchior2018}, \citealt{Nourbakhsh2021}, and \citealt{Melchior2021} for a review). The inaccuracies due to blending are alarming for many of the next generation of large cosmology surveys where precision measurements of the expansion history of the Universe as well as the growth rate of large scale structure, heavily rely on photometric data sets. 

\begin{figure*} 
\centering
  \includegraphics[trim=0cm 0cm 0cm 0cm, clip,width=1.0 \textwidth] {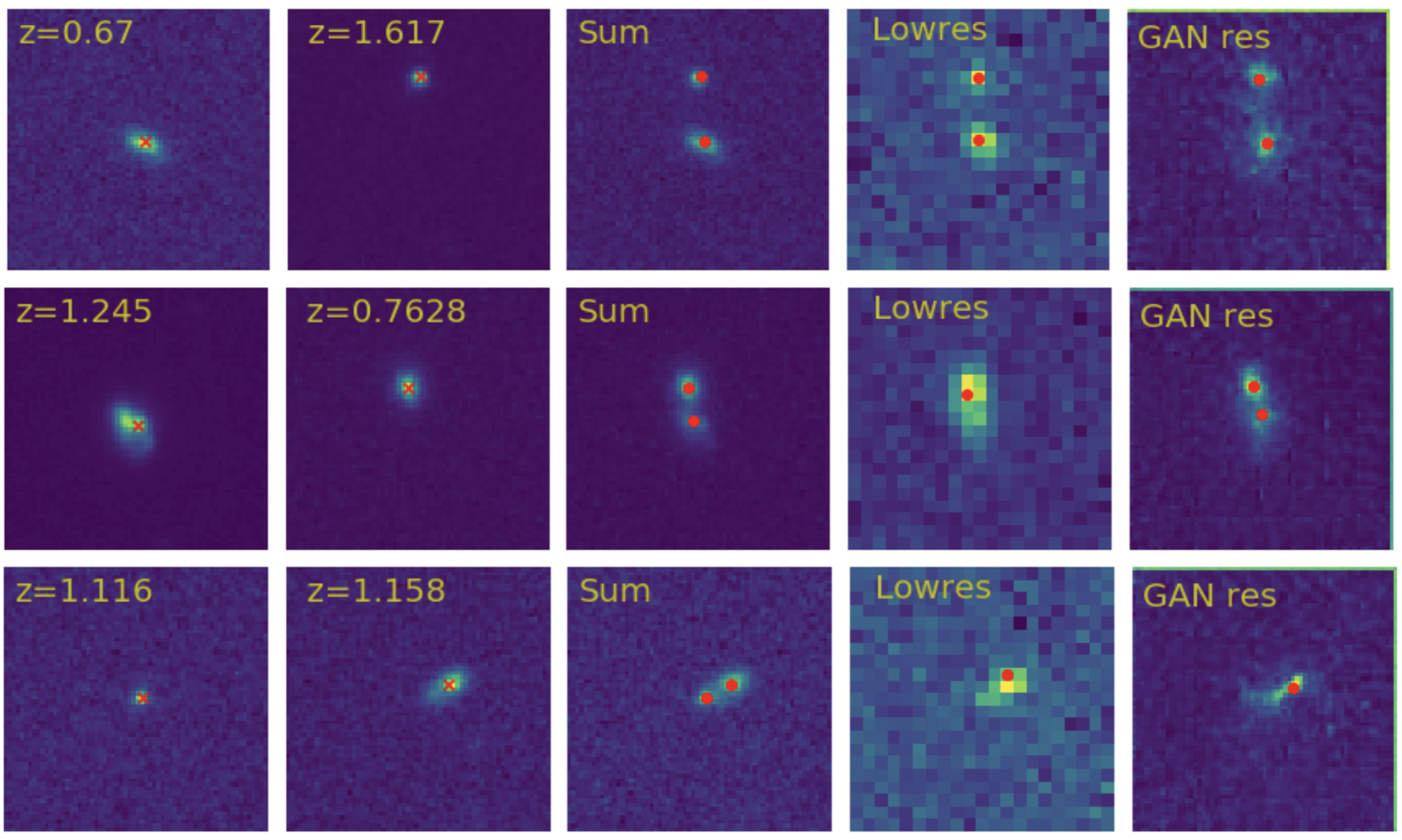}
\caption{Peak finder on different resolutions of simulated blend sample. From left to right, first and second column are randomly drawn HST F775W cutouts of two galaxies moved to a random off-center position and orientation. Third column adds the first two to create the blended sample in high resolution. Downgraded image which is fed to GAN is shown on the fourth column. Fifth column shows GAN output improving the resolution of the downgraded image. Crosses on the first two columns correspond to main sources we want to detect. Red dots on the last three columns show peaks found in each cutout.}
\label{fig:peakfinder}
\end{figure*}

We check the performance of this GAN architecture in deblending galaxies as a function of the properties of the two galaxies in the blend (i.e., angular distance, flux ratio, redshift difference, and sizes). We construct a sample of five thousand blended cutouts. Figure \ref{fig:peakfinder} shows three examples of our constructed blends. The two left columns show two galaxies in the blend marked with a red cross. Third, fourth and fifth columns show the sum of the first two in high resolution, degraded to lower resolution, and recovered resolution with the GAN. Red circles represent the peaks found by a peak finder algorithm which takes the PSFs into account. We position match the peaks with the initial galaxies in the blend. The main cases where our GAN helps are those where the two sources are detected in the high-resolution image, only one in the low resolution (so-called ambiguous or undetected blends in \citealt{Dawson2016}; as shown in the middle row of Figure \ref{fig:peakfinder}) and again both are detected in the recovered one. We measure a false positive rate of $\sim 1\%$ by testing this network on a sample with no blends as well as cases in the blended sample where there is only one peak in the high and low resolution and more than one in the GAN resolution.

\begin{figure*} 
\centering
  \includegraphics[trim=0cm 0cm 0cm 0cm, clip,width=0.99 \textwidth] {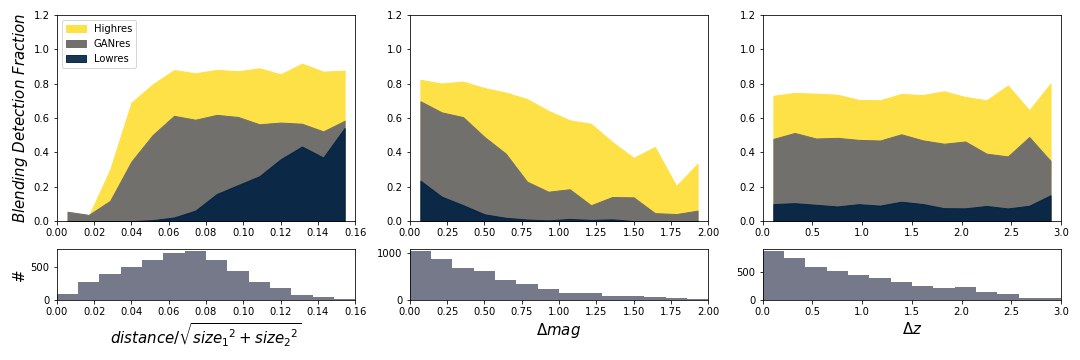}
\caption{Performance of GAN in deblending as a function of distance and sizes of the two objects as well as their magnitude and redshift difference (respectively from left to right). Yellow, grey and dark blue shaded regions corresponds to the fraction of successfully finding two peaks in the high, GAN, and low resolution cutout, respectively. The histograms on the bottom row show the number of simulated blends at each bin.}
\label{fig:fractions}
\end{figure*}

Performance of our method in recovering the two objects in the blended cutouts is shown in Figure \ref{fig:fractions}. We define the blend detection fraction as the fraction of cases where the main two peaks are found in the cutout and study its variation for the high, low, and GAN resolutions. The overall trends with the angular distance divided by the sum of the sizes and flux ratio of the two objects show what one would expect, namely the fraction increasing with the increase in the angular distance and decreasing with the increase in the flux ratio. The smaller scale ups and downs, specifically towards larger distances, and brightness/redshift differences are less significant and vary with a change in binning given the lower number of objects in those bins as shown in the bottom panels. The main take away from the first two subplots is the boost in the fraction using the GAN compared to the low resolution. Naturally, even with HST resolution if the angular distance between the two objects is too small a large fraction of blends would be missed. However, at larger angular distances ($\sim 0.8\arcsec$) where most of the blends are detected in the high resolution and many missed in the lower resolution, GAN boosted resolution improves significantly in the detection fraction. We divide the distances by the square root of the sum of the sizes of each of the objects in the blend. We use the FWHM reported in the GOODS south catalog which are measured by SExtractor on F160W images. This is an approximate measure of the size of the galaxy and not very precise due to both being in a different waveband as well as being measured by SExtractor which is not fine tuned for measuring sizes. The trends seen in the left panel of Figure \ref{fig:fractions} are mostly due to angular distance and we see no significant trend with sizes. 

At very low flux ratios where around 80$\%$ of blends are detected in the high resolution and only $\sim 20\%$ in the low resolution image, with the GAN improved resolution $\sim 60 \%$ of the blends can be detected. This is not the case at higher flux ratios where no improvement in detection is gained by using the GAN as one of the objects outshines the other one. 

No significant change in the blend detection fraction as a function of redshift difference between the two sources is found (right panel of Figure \ref{fig:fractions}). This shows the overall performance of the peak finder in the simulated blends to be $\sim 80, 10,50 \%$ respectively in the high, low, and GAN resolution cutouts. 

The analysis presented so far demonstrates improvements in the detection of blended objects in images with GAN enhanced resolutions. To better visualize this improvement as a function of the variables in the simulated blends, we use Self Organizing Maps (SOMs; \citealt{kohonen}) and reduce the dimensions of the parameter space. SOMs are a class of unsupervised neural networks that reduce the dimensions of the multi-dimensional space and map the manifold into a two dimensional space while preserving the topology. SOMs have been vastly used recently in the astronomy literature, mostly in tuning redshifts for cosmology projects (e.g., \citealt{Masters2017}, \citealt{Hemmati2019}), for measuring the physical properties of galaxies (e.g., \citealt{Hemmati2019-2}, \citealt{Davidzon2019}), as well as in classifying modified gravity theories (\citealt{Ferte2021}). Here we train a 32x24 SOM grid on the distances, flux ratio, sum of the sizes and redshift difference of the sample of five thousand blended galaxies. The shape and size of the SOM grid are not strict and are only chosen in order to minimize the dispersion of data within a given cell, but at the same time to avoid a large fraction of them to be empty. The median and mean number counts are $\sim 6$ per cell with a dispersion of $\sim 3$.

\begin{figure*} 
\centering
  \includegraphics[trim=0cm 0cm 0cm 0cm, clip,width=0.99 \textwidth] {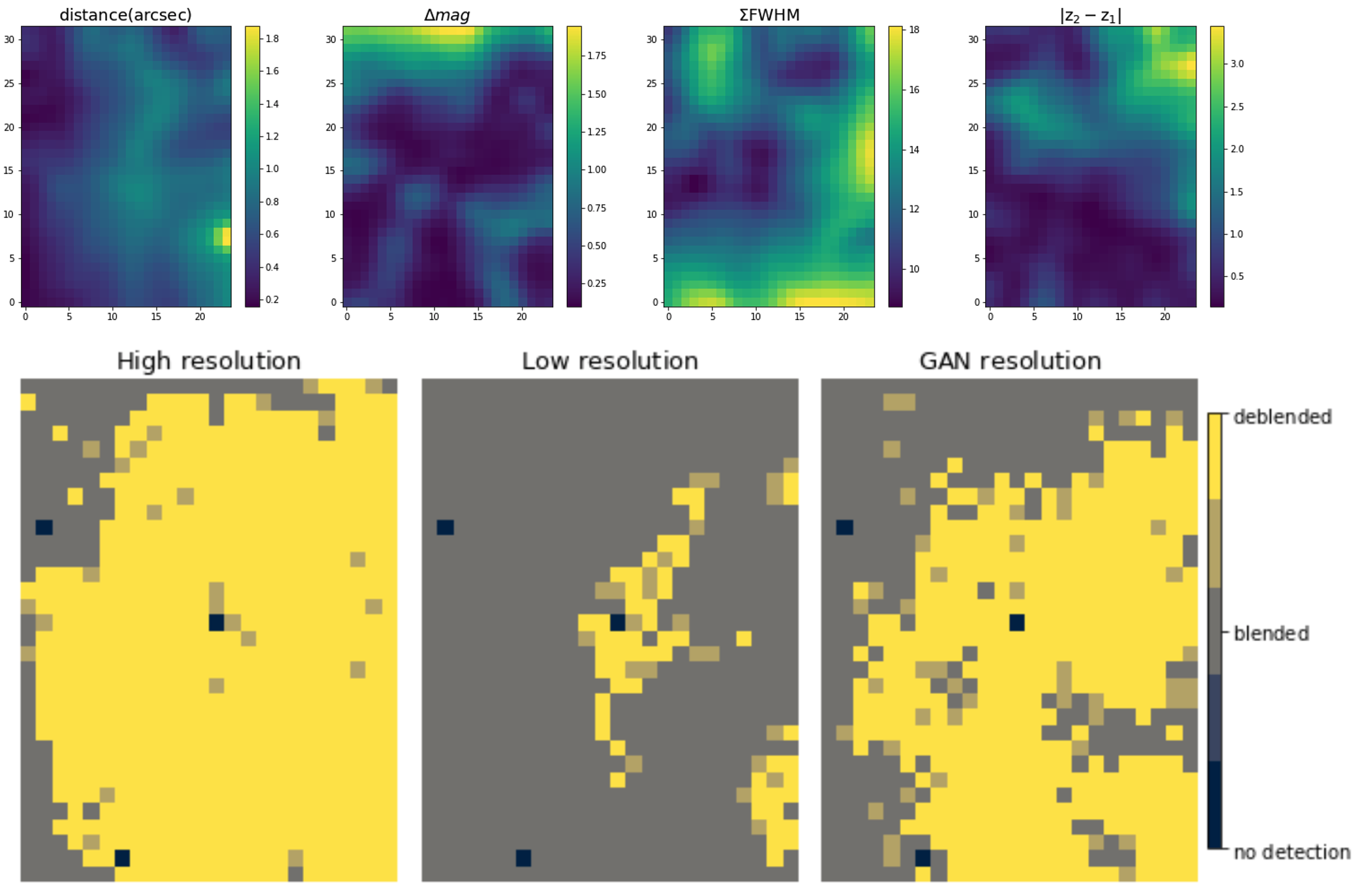}
\caption{A Self Organizing Map trained on the distance, flux ratio, $\Sigma FWHM$, and $\Delta z$ of five thousand simulated blended sample. The 32x24 grids on the first row show how the four dimensional parameter space projects onto two dimensions. Second row shows successful peaks found as a function of position on the two dimensional parameter space in high, low and GAN resolution images (left to right). Yellow, gray and dark blue respectively are where two peaks were found (deblend successful), one peak was found (blended), and where no object was detected.}
\label{fig:SOM}
\end{figure*}
\begin{figure*}
\centering
  \includegraphics[trim=0cm 0cm 0cm 0cm, clip,width=0.99 \textwidth] {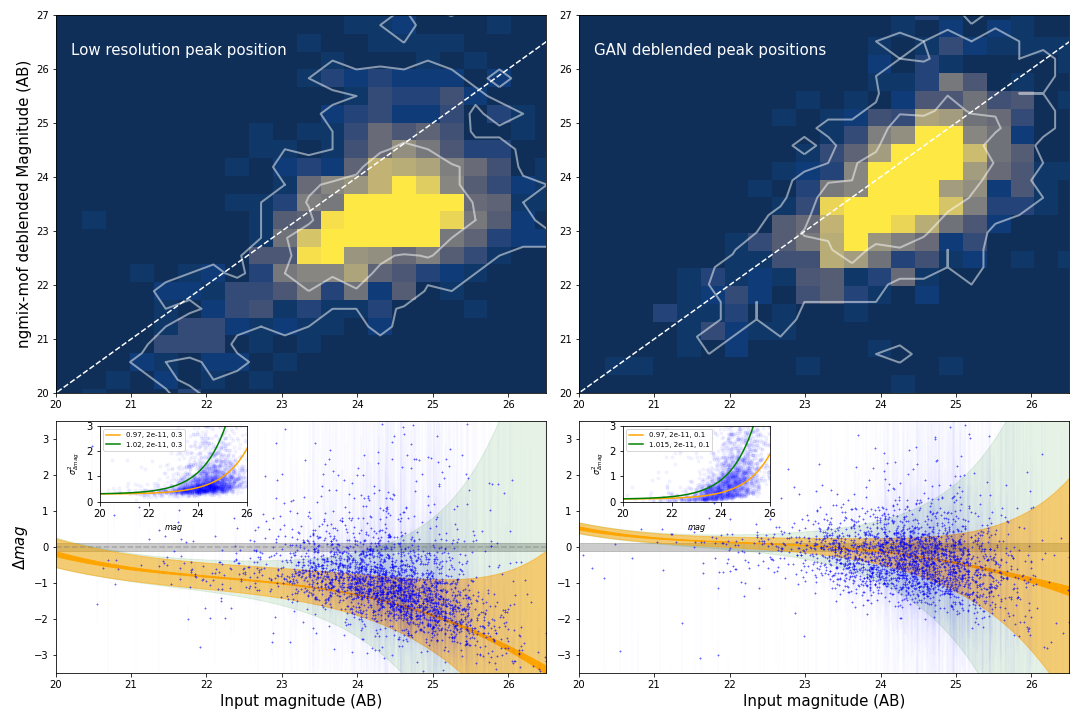}
\caption{Photometry of sources with and without GAN's deblended positions on the right and left panels respectively. Contours show 1 and 2 $\sigma$ distribution and the dashed white line shows the 1:1. Bottom panel plots the magnitude difference between the input and measured photometry as a function of the input magnitude with the Gaussian Processes Regression (GPR) over plotted (see section 2.2 of \cite{RW} for detailed equations of GPR). Darker orange shaded region corresponds to the predictive variance around the predictive mean assuming noise free data. The brighter orange and green shaded regions correspond to two levels of noise variance from the data points added to the predictive variance. The inset plots show the variance of $\Delta magnitude$ plotted as a function of magnitude with exponential functions of the form $y=a*exp(bx)+c$ over plotted corresponding to the median (orange) and including 68$\%$ of the data points (green).}
\label{fig:photometry}
\end{figure*}

Figure \ref{fig:SOM} presents this SOM. The four panels in the top row show the four dimensional input vector spread over the grid. Each simulated blend is mapped to the SOM and cells are colored with the average parameter (i.e., angular distance, magnitude difference, $\Sigma$ FWHM, and redshift difference) of the objects in each cell. Different locations on the two dimensional grid hence correspond to a unique combination in the four dimensional input space, distance between the two galaxies in each blend for example mostly increases from left to right of the grid and the highest flux ratios are mostly concentrated on top of the grid. On the second row we color the grid with the average blend detection fraction in the high, low and GAN resolution images from left to right. Dark red represents cells where two galaxies are detected by the peak finder, pink cells are those where both cases exist some one and some two peaks detected in the cutouts, gray cells are those where the two objects are fully blended and white cells have no cutouts mapped to them. Segments of the full parameter space where blending is an issue and where our GAN architecture helps resolve is clearly disclosed with SOMs visualization. We note that this figure represents the manifold constructed based on the parameter space assumed in our simulation. For instance if one models mostly bright galaxies and large angular separations a larger portion of the low resolution grid could be filled.

\subsection{Photometry improvement}

With the ability of GANs in detecting a large fraction of objects blended in low resolution images, we look into the improvement in the photometry measurements given the recovered coordinates. To do photometry, we use the recovered coordinates from GAN on the low resolution images rather than using the whole GAN constructed image. This is more precise as the fluxes are not preserved in the reconstruction due to scaling of images in the training. It also reduces the uncertainty of shapes, backgrounds, noise, etc. changing in the process. 

We use the NGMIX (\citealt{Sheldon2015}, \citealt{Sheldon2020}) software which is used for shear measurements in weak lensing studies to do photometry. In NGMIX, both the PSF profile and the galaxy are modelled using mixtures of Gaussians and convolutions are performed analytically, resulting in fast model generation as compared to methods that perform the convolution in Fourier space (\citealt{Jarvis2016}). Figure \ref{fig:photometry} shows the improvement in photometry measurement given the more accurate positions found by the GAN. The plot shows the measured vs. the input photometry of the dominant source in the cutouts. The input magnitudes shown on the x axis are measured from the high resolution image, PSF and peak positions using NGMIX. And the measured ones on the y axis are performed on the low resolution image and PSF with the peak position from the low resolution image as well as the two positions from the GAN images respectively on the left and right panels. The improvement in photometry measurement is evident from the shift of the distribution and contour plots upwards toward the 1:1 line. In the bottom subplots of Figure \ref{fig:photometry} we show the magnitude difference as a function of input magnitudes. We account for photometry errors and fit using a Gaussian Process (GP) Regression (\citealt{Williams96}) of the scikit-learn package (\citealt{scikit-learn}) which is a non parametric Bayesian approach to regression. The GP kernel adopted here is the composition of the constant kernel with the radial basis function kernel. More than an order of magnitude improvement in the flux measurement is achieved using the GAN deblender. 

\begin{figure*}
\centering
  \includegraphics[trim=0cm 0cm 0cm 0cm, clip,width=0.99 \textwidth] {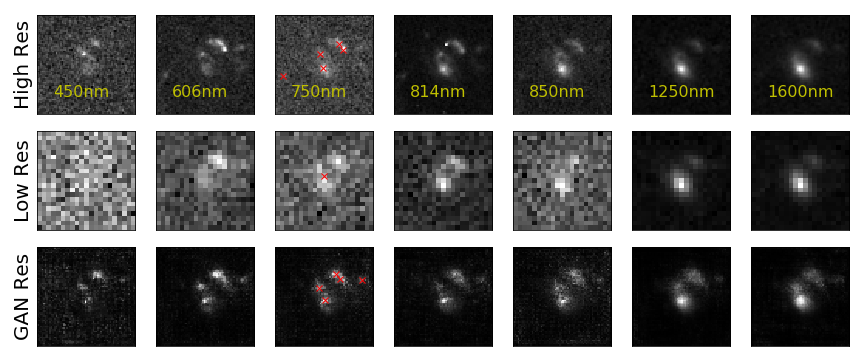}
\caption{GAN trained on multi-wavelength HST cutouts. From left to right columns correspond to F435W, F606W, F750W, F814W, F850lp, F120W, F160W. Rows show the HST, Subaru and GAN recovered resolutions from top to bottom. Peak detection using the PSFs is shown by red crosses on the iband cutouts in each of the resolutions.}
\label{fig:multi}
\end{figure*}


\subsection{Multi-wavelength GAN}

\begin{figure}
\centering
  \includegraphics[trim=0cm 0cm 0cm 0cm, clip, width=0.49 \textwidth] {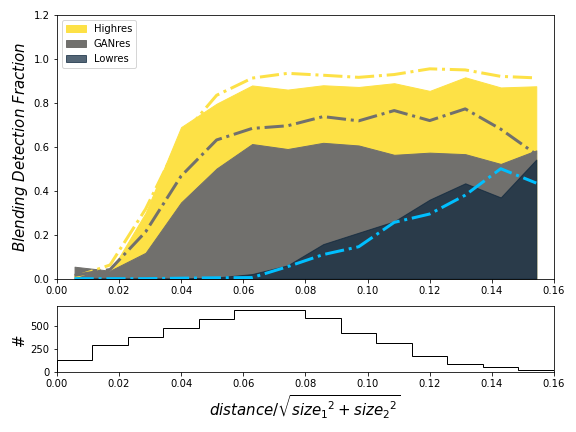}
\caption{Same as the left panel of Figure \ref{fig:fractions}. Dash-dotted lines show the improved 
fractions measured using the multi-wavelength GAN. The fall in the list bin is insignificance and due to both poor statistics and proximity to edges of the cutouts.}
\label{fig:multimprove}
\end{figure}

The results presented above can be further improved training a multi-channel GAN. We test this using seven band high resolution HST (F435W, F606W, F775W, F814W, F850lp, F125W, F160W) images and downgrading them to the Subaru resolution as explained above. We similarly constructed a sample of 5000 blended cutouts in multi-bands. The multi-wavelength GAN has a modified architecture in the sense that in the input as well as the PyTorch conv2d layers the number of channels is changed from 1 to seven. We tuned the rest of the hyper-parameters (e.g., padding, stride, bias, etc.) accordingly to have the right output dimensions (Figure \ref{fig:architecture} shows the architecture of the multi-channel GAN in the bottom panel).

Figure \ref{fig:multi} shows one example where the rows correspond to the high, low and GAN resolution respectively and the columns correspond to different wavebands with increasing wavelength from left to right. Crosses on the F775W (i band) show detected sources in different resolutions given the corresponding PSFs. Similar results to a single band GAN is seen with a multi-wavelength GAN. However, the GAN trained on multi-wavelength SED is favorable for detections as it also contains information from longer and shorter wavelengths. Figure \ref{fig:multimprove} shows an additional 10 $\%$ improved fraction of successful blending detection with the multi-wavelength GAN. The same photometry improvement as single band is seen on the successful deblended sample (as expected), by performing the peak finding on the i-band cutouts. While Figure \ref{fig:multi} shows that this GAN architecture correctly recognizes the morphological variation with wavelength it also makes it clear why we mainly use GANs for detection of peak positions rather than photometry as the fluxes are not preserved in the recovered images (see e.g., F450w where nothing is detectable in the low resolution and clear detections exist in the GAN res). 

\section{Discussion}\label{sec:disc}

We showed that simple GAN architectures can be used to generate galaxies with HST resolution either starting from a noise vector or by increasing the resolution of ground-based images. We demonstrated the appropriateness of GANs in detection of blended galaxies in low resolution images and the gain in the photometry of blended objects either using only a single band or the optical+NIR SED. We note that while GANs are practical tools with superb performance in many settings, they are only good approximations of the underlying probability distribution of the data and do not guarantee to learn and represent the exact distribution. They are also not the only viable deep neural network solution to handle blending. For instance, the reduction of undetected blends has been done also by e.g. Kamath (2020, PhD thesis) with a Mask-RCNN architecture. \citealt{Boucaud2020} used a U-Net architecture (\citealt{unet}- a CNN which passes data through a bottleneck) to simultaneously find the blends and their segmentation maps in monochrome galaxy cutouts. \citealt{Arcelin2020} used a combination of two VAEs; one for denoising isolated galaxies and used it as a prior in another VAE like network for deblending. Also using an iterative combination of two networks; a residual CNN which isolates a single galaxy from a composite cutout and a classifier which counts the remaining peaks in the cutout; \citealt{Wang2022} constructed a deblender which outperforms the industry standard, SExtractor (\citealt{Sextractor}). A compelling variant of GANs is also introduced and tested on nearby SDSS galaxies by \citealt{Reiman2019}, where they have the generator output two cutouts given one composite input cutout and fine tune it by giving these two deblended cutouts as well as the two that were used to construct the composite to the discriminator. While in this work, we improved the resolution/noise/pixel scale of ground-based images by training from the space data using GANs and did source detection on the enhanced products using a standard peak-finding routine, \citealt{Buchanan2022} showed that detection of blends can be further improved using either Gaussian Process or CNN-based blend classifiers. They also showed that the improvement in blend identification depends strongly on the footprint image they start with. Hence, a combination of an image enhancement network (e.g., GAN based) and an optimized classifier (e.g., GP based) seems to be advantageous over non DL methods and over either one alone for deblending galaxies. In future work, we will also explore alternative neural network structures for resolution boosting and denoising of astronomical images including the deep image prior (\citealt{Ulyanov2017}) and diffusion probabilistic models (\citealt{Saharia2021}).

We simulated the blend sample from real galaxy cutouts, as opposed to using Galsim (\citealt{galsim}) models based on real galaxy distributions (e.g., \citealt{Arcelin2020}, \citealt{Buchanan2022}, \citealt{Wang2022}) to be more representative of real galaxies. We restricted the training sample to galaxies brighter than the 24th magnitude in F750W and in testing went down to 25th magnitude. In future work we will explore the possibility to push this boundary deeper, which is possible given the depth of HST images in the CANDELS fields. We also focused here mainly on cases where two galaxies are blended. This can be improved upon by having simulations of different fractions of blends of multiple galaxies as expected from deep surveys (e.g., the Blending Toolkit from the DESC project, \citealt{Sanchez2021}). While the redshifts, sizes and magnitudes of the galaxies in the blends are drawn from the CANDELS catalogs and hence representative of real populations, angular distances are drawn from a normal distribution. This choice affects the fraction of blended objects in our simulations. Here, we only explored the improvements in blending detection and photometry going from HSC to HST resolutions. Further work is needed to explore the parameter space of low resolution images as well as recovery of the backgrounds to see the extent of applicability of GANs. 

A large fraction of galaxies ($\sim 58\% $) are blended in HSC or LSST depths (\citealt{Bosch2018}, \citealt{Sanchez2021}), hence deblending is crucial for precision cosmology. Data from space such as those from the Roman or Euclid suffer less from blending (e.g., \citealt{Chary2020}). However they still rely on ground-based data such as those from the LSST to have accurate photometric redshifts (e.g., \citealt{Hemmati2019}). The multi-wavelength trained GAN also opens a potential avenue for the SPHEREx mission (\citealt{Bock2018}) to improve the photometry as well as redshifts of $\sim 1.4$ billion galaxies (e.g., \citealt{Cooray2018}) with low spectral ($R \sim 40 -130$) and poor spatial (6" pixels) resolution. It remains to be quantified how much the deep learning image enhancement adds to the performance of current forced-photometry techniques (e.g., \citealt{Laider2007P}, \citealt{Lang2016}) using external and higher resolution data, especially by being applied where the high resolution data is missing. 

We wish to thank the referee for constructive comments which greatly improved the content and presentation of this paper. S.H is thankful to A. Cooray group at UCI who let us use their GPU servers. We are thankful to NVIDIA for the GPU granted as their academic grant program. This work used {\sc SOMPY}, a python package for self organizing maps (main contributers: Vahid Moosavi @sevamoo, Sebastian Packmann @sebastiandev, Iv\'an Vall\'as @ivallesp). Parts of this research were carried out at the Jet Propulsion Laboratory, California Institute of Technology, under a contract with the National Aeronautics and Space Administration. And parts, by IPAC at the California Institute of Technology, and was sponsored by the National Aeronautics and Space Administration. This work was in part funded by the HST cycle 29 archival program, 16615.

\bibliography{gan.bib}

\end{document}